\documentclass[10pt,twocolumn]{article}

\usepackage{latex8}

\usepackage{times}
\usepackage{graphicx}

\newif\ifremark
\long\def\remark#1{
\ifremark%
    \begingroup%
    \dimen0=\textwidth
    \advance\dimen0 by -1in%
    \setbox0=\hbox{\parbox[b]{\dimen0}{\protect\em #1}}
    \dimen1=\ht0\advance\dimen1 by 2pt%
    \dimen2=\dp0\advance\dimen2 by 2pt%
    \vskip 0.25pt%
    \hbox to \textwidth{%
        \vrule height\dimen1 width 3pt depth\dimen2%
        \hss\copy0\hss%
        \vrule height\dimen1 width 3pt depth\dimen2%
    }%
    \endgroup%
\fi} \remarktrue
\begin{document}

\title{
{\bf Failure Data Analysis of HPC Systems}
   \thanks{
    This work was supported in part by
    Contract 74837-001-0349 from the Regents of University of
    California (Los Alamos National Laboratory) to William Marsh
    Rice University and by National Science
    Foundation under grant ACI 02-19597.
   }
}

\author{
Charng-Da Lu\\
Buffalo, NY 14214
}

\maketitle
\thispagestyle{empty}

\begin{abstract}

Continuous availability of HPC
systems built from commodity components have become a
primary concern as system size grows to thousands of 
processors. In this paper, we present the analysis of
8-24 months of real failure data collected from three 
HPC systems at the National Center for Supercomputing 
Applications (NCSA). The results show that the availability 
is 98.7-99.8\% and most outages are due to software halts. 
On the other hand, the downtime are mostly contributed by 
hardware halts or scheduled maintenance. We also used
failure clustering analysis to identify several correlated 
failures. 

\end{abstract}

\Section{Introduction}
Continuous availability of high performance computing (HPC) 
systems built from commodity components have become a primary 
concern as system size grows 
to thousands of processors. To design more reliable systems, 
a solid understanding of failure behavior of current systems
is in need. Therefore, we believe failure data analysis of 
HPC systems can serve three purposes. First, it highlights
dependability bottlenecks and serves as a guideline for
designing more reliable systems. Second, real data can 
be used to drive numerical evaluation of performability
models and simulations, which are an essential part of
reliability engineering. Third, it can be applied to predict
node availability, which is useful for resource characterization
and scheduling \cite{Brevik:03}.

In this paper, we studied 8-24 months of real failure data
collected from three HPC systems\footnote{All NCSA HPC 
systems described in this paper have been decommissioned 
in 2003 and 2004.} at the National Center for 
Supercomputing Applications (NCSA). 
The remainder of this paper is organized as follows. In
\S\ref{s:sys} we described the systems characteristics
and failure data collection. We present preliminary analysis of
failure data in \S\ref{s:prelim}, followed by failure 
distribution and correlation analysis in \S\ref{s:dist} 
and \S\ref{s:corr}. We summarize related work in
\S\ref{s:rel} and conclude our study in 
\S\ref{s:concl}.

\Section{The Systems and Measurements}
\label{s:sys}
The three HPC systems we studied are quite different 
architecturally. The first is an array of SGI Origin 
2000 (O2K) machines. 
SGI Origin 2000 is a cc-NUMA distributed 
shared memory supercomputer. An O2K can have up to 512
CPUs and 1 TB of memory, all under control of 
one single-system-image IRIX operating system. 
The configuration at NCSA is an array of twelve O2K's
(total 1520 CPUs) connected by proprietary, high-speed HIPPI 
switches. Table~\ref{tbl:failuredataO2K} lists its detailed 
specification. The machines A, B, E, F, and N are equipped 
with 250 MHz MIPS R10000 processors, and the rest with 
195 MHz MIPS R10000 processors. M4 accepts interactive 
access, while the others machines only service 
batch jobs. Peak performance of NCSA O2K is 
328 gigaflops. 

The second and the third systems are Beowulf-style PC clusters.
``Platinum'' cluster has 520 two-way SMP 1 GHz Pentium-III nodes 
(1040 CPUs), 512 of which are compute nodes (2 GB memory), 
and the rest are storage nodes and interactive access nodes 
(1.5 GB memory). ``Titan'' cluster consists of 162 two-way 
SMP 800 MHz Itanium-1 nodes (324 CPUs), 160 of 
which are compute nodes (1.5 GB memory) and 2 are for 
interactive access. Both clusters use
Myrinet 2000 and Gigabit Ethernet as system interconnect. 
Myrinet is faster and for node communications, whereas 
the Gigabit Ethernet is slower and serves I/O traffic.
Both clusters have one teraflop of peak performance. 

All three HPC systems use batch job control software to manage
workload. O2K runs LSF (Load Sharing Facility) queueing system.
Each job on O2K have resource limits of 50 hours of run-time
and 256 CPUs. Platinum and Titan employ Portable Batch 
System with the Maui Scheduler, and the job limits are 352
and 128 nodes for 24 hours, respectively.

According to a user survey \cite{NCSASurvey}, the NCSA HPC 
systems are devoted to multiple disciplinary sciences research: 
physics (20\%), engineering (16\%), chemistry (14\%), 
biology (13\%), astronomy (13\%), and material science (12\%). 
Seventy percent of users write programs in Fortran (F90 and F77) or mix of 
Fortran and C/C++. Sixty-five percent users use MPI or 
OpenMP as the parallel programming model. In terms of job 
sizes, 22\% users typically allocate 9-16 CPUs. About equally 
many users (14-15\%) allocate 2-4, 5-8, 17-32, or 33-64 CPUs.

The failure log was collected in the form of monthly or
quarterly reliability reports. At the end of a month or aquarter, 
a report for each node/machine is created. A report records outage
date (but no outage time), type, and duration. There are 
five outage types defined by NCSA system administrator: 
Software Halt (SW), Hardware Halt (HW), Scheduled Maintenance (M), 
Network Outages, and Air Conditioning or Power Halts (PWR). 
The cause of an outage is determined as follows: a program runs 
at machine boot time prompts the administrator to enter the 
reason for the outage. If nothing is entered after two 
minutes, the program defaults to recording a Software Halt.

The data collection period was two years (April 2000 to March 2002)
for O2K and eight months (January 2003 to August 2003) for 
Platinum and Titan. In this set of failure log, there is no 
occurrence of Network Outage, so we exclude it from the rest of 
analysis.

\begin{table*}

\small{

\begin{tabular}
{|p{49pt}|p{18pt}||p{18pt}|p{18pt}|p{18pt}|p{18pt}|p{18pt}|p{18pt}|p{18pt}|p{18pt}|p{18pt}|p{18pt}|p{18pt}|p{18pt}|}
\hline
& 
All& 
A& 
B& 
E& 
F& 
H1& 
H& 
J& 
M& 
M2& 
M4& 
N& 
S \\
\hline
CPUs& 
1520& 
128& 
256& 
128& 
128& 
128& 
128& 
128& 
128& 
64& 
48& 
128& 
128 \\
\hline
Mem (GB)& 
618 &
64& 
128& 
64& 
76& 
64& 
32& 
32& 
32& 
16& 
14& 
64& 
32 \\
\hline
\hline
Outages& 
687& 
87& 
182& 
40& 
81& 
42& 
25& 
32& 
24& 
41& 
59& 
37& 
37 \\
SW ({\%})& 
59& 
74& 
68& 
53& 
63& 
57& 
60& 
59& 
42& 
44& 
39& 
49& 
46 \\
HW({\%})& 
13& 
8& 
19& 
13& 
9& 
21& 
8& 
3& 
17& 
10& 
5& 
5& 
32 \\
M({\%})& 
21& 
11& 
12& 
28& 
20& 
12& 
24& 
19& 
29& 
32& 
49& 
38& 
14 \\
PWR({\%})& 
7& 
7& 
2& 
8& 
9& 
10& 
8& 
19& 
13& 
15& 
7& 
8& 
8 \\
\hline
Downtime \par (day)& 
9.5& 
8.7& 
19.2& 
15.3& 
5.9& 
13.8& 
6.2& 
3.6& 
4.8& 
7.5& 
4.5& 
5.5& 
5.0 \\
SW ({\%})& 
27& 
32& 
49& 
11& 
36& 
9& 
5& 
25& 
15& 
27& 
12& 
6& 
16 \\
HW({\%})& 
28& 
35& 
29& 
1& 
10& 
67& 
49& 
$<1$& 
22& 
12& 
4& 
28& 
22 \\
M({\%})& 
41& 
28& 
20& 
85& 
44& 
22& 
45& 
66& 
58& 
52& 
75& 
62& 
58 \\
PWR({\%})& 
4& 
4& 
2& 
3& 
9& 
2& 
2& 
9& 
4& 
9& 
10& 
4& 
4 \\
\hline
Avail({\%})& 
98.7& 
98.8& 
97.4& 
97.9& 
99.2& 
98.1& 
99.2& 
99.5& 
99.3& 
99.0& 
99.4& 
99.2& 
99.3 \\
\hline
S Avail({\%})& 
99.2& 
99.1& 
97.9& 
99.7& 
99.6& 
98.5& 
99.5& 
99.8& 
99.7& 
99.5& 
99.8& 
99.7& 
99.7 \\
\hline
MTBF (day)& 
1.0&
8.1&
4.0&
15.9&
8.6 &
14.7 &
29.5 &
22.5 &
30.3 &
17.4 &
12.3 &
18.6 &
18.5 \\
StdDev& 
2.1& 
14.5& 
5.8& 
28.5& 
14.7& 
20.9& 
36.9& 
33.4& 
48.3& 
31.7& 
18.7& 
34.3& 
22.9 \\
Median & 
0.9& 
1.7& 
2.1& 
1.7& 
2.5& 
3.5& 
25.0& 
1.0& 
4.0& 
1.6& 
5.7& 
1.0& 
11.0 \\
\hline
MTTR (hr)& 
3.5& 
2.4& 
2.5& 
9.2& 
1.7& 
7.9& 
6.0& 
2.7& 
4.8& 
4.4& 
1.9& 
3.6& 
3.2 \\
StdDev& 
13.1& 
7.8& 
5.2& 
29.9& 
5.1& 
33.2& 
15.6& 
7.5& 
9.3& 
7.7& 
8.1& 
8.8& 
7.2 \\
Median & 
0.5& 
0.4& 
0.9& 
0.43& 
0.4& 
0.5& 
0.5& 
0.3& 
0.5& 
0.7& 
0.4& 
0.4& 
0.5 \\
\hline
MTTR SW& 
1.5& 
1.1& 
1.8& 
1.9& 
1.0& 
1.3& 
0.5& 
1.1& 
1.7& 
2.7& 
0.6& 
0.4& 
1.1 \\
MTTR HW& 
6.3& 
10.5& 
3.9& 
0.6& 
2.0& 
24.7& 
36.3& 
0.4& 
6.4& 
5.4& 
1.3& 
18.6& 
2.2 \\
MTTR M& 
8.0& 
5.8& 
4.3& 
28.6& 
3.9& 
14.6& 
11.2& 
9.6& 
9.6& 
7.1& 
2.8& 
5.9& 
13.8 \\
MTTR PWR& 
2.1& 
1.5& 
3.4& 
3.8& 
1.0& 
1.5& 
1.3& 
1.2& 
1.7& 
2.8& 
2.6& 
1.7& 
1.7 \\
\hline
\end{tabular}

}

\caption{O2K Failure Data Summary}
\label{tbl:failuredataO2K}
\end{table*}

\begin{table*}
\begin{center}

\small{

\begin{tabular}{rl}

 \begin{tabular}{|l|r|r|}
 \hline
              & Platinum   & Titan  \\
 \hline
 Outages      &  7279     &  947   \\
 Outage/Node  &   14.00  &   5.85 \\
 SW (\%)  &  84       &  60   \\
 HW (\%)  &  $<0.1$   &   5   \\
 M  (\%)  &  16       &   1   \\
 PWR (\%) &  0        &  34   \\
 \hline
 Downtime/Node (hr) &  12.16 &   12.55   \\
 SW (\%)  &  69       &  18   \\
 HW (\%)  &  10       &  18   \\
 M  (\%)  &  21       &  $<1$ \\
 PWR (\%) &  0        &  64   \\
 \hline
 Avail (\%)    &  99.79 & 99.78 \\
 \hline
 S Avail (\%)  &  99.83 & 99.79 \\
 \hline
 \end{tabular}

&

 \begin{tabular}{|l|r|r|}
 \hline
              & Platinum   & Titan  \\
 \hline
 System MTBF (hr) &  0.79   &  5.99 \\
 StdDev      &  5.77       &  40.09 \\
 \hline
 Node  MTBF (day)  &  14.16  &  26.89 \\
 StdDev      & 15.87       &  25.75 \\
 Median      &  9          &   29   \\
 \hline
 Node MTTR (hr) &  0.87  & 2.15 \\
 StdDev      &  4.27       & 4.65  \\
 Median      &  0.15       & 0.28  \\
 \hline
 MTTR SW     &  0.70     & 0.63  \\
 MTTR HW     &  100.67   & 7.60  \\
 MTTR M      &  1.15     & 0.55  \\
 MTTR PWR    &  $-$      & 4.08  \\
 \hline
 \end{tabular}

\end{tabular}

}

\caption{Platinum and Titan Failure Data Summary}
\label{tbl:failuredataPT}
\end{center}
\end{table*}

\begin{figure}[h]
   \begin{center}
   \includegraphics[scale=0.65]{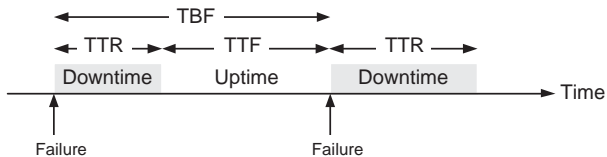} 
    \caption{TBF, TTF, and TTR}
    \label{fig:TBFTTFdiff}
   \end{center}
\end{figure}

\Section{Preliminary Results}
\label{s:prelim}
Before describing the failure data, we would like to clarify some 
terminology. Time to Failure (TTF) is the interval between the 
end of last failure and the beginning of next failure.  
Time between Failures (TBF) is the interval between the 
beginnings of two consecutive failures. Time to Repair (TTR) 
is synonymous with Downtime. Figure~\ref{fig:TBFTTFdiff} 
illustrates the differences. Because the failure log does not 
include the start and end times of outages, we can only 
calculate TBFs in terms of days.

\begin{figure*}
   \begin{center}
   \includegraphics{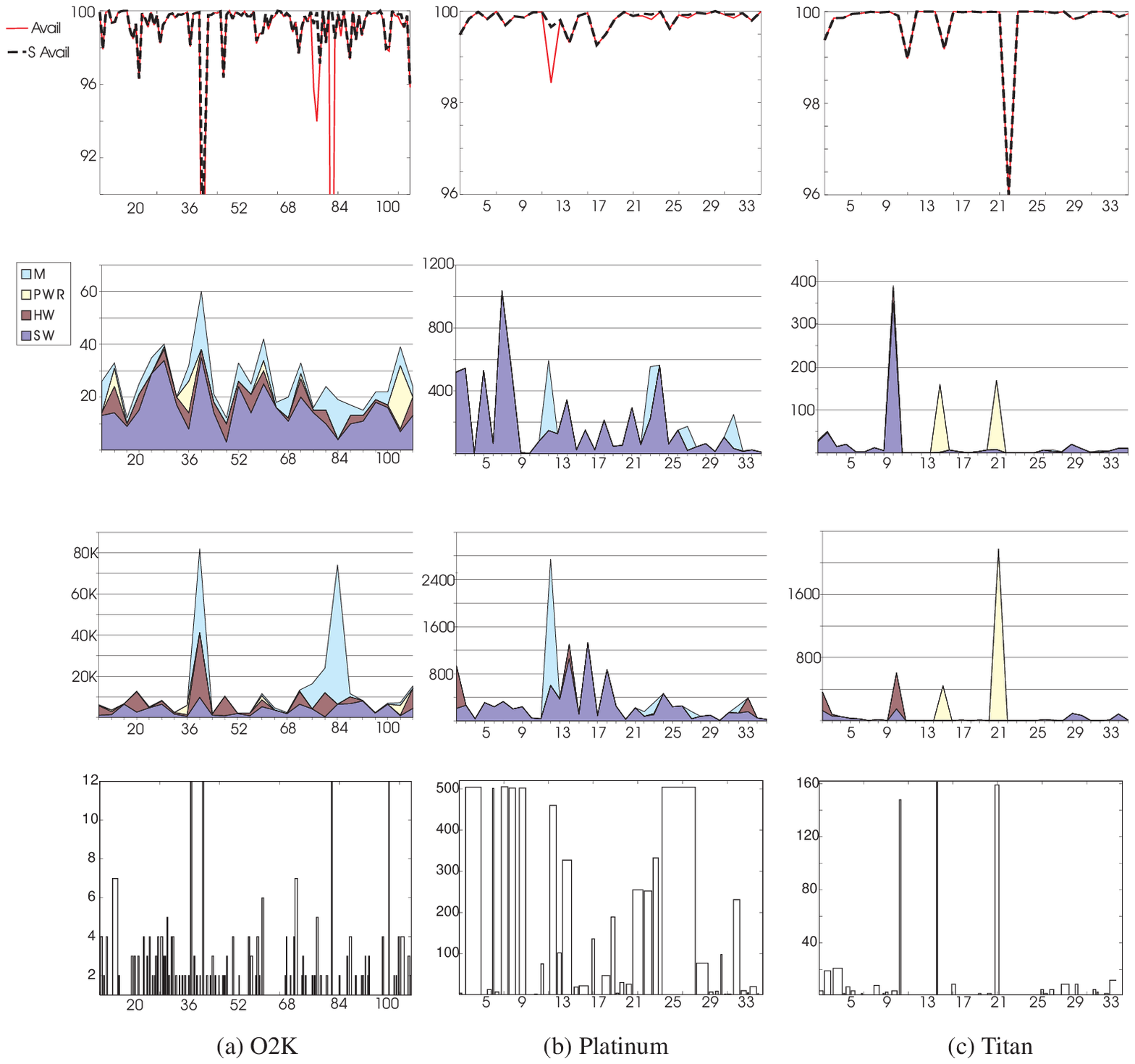} 
    \caption{The rows from top to bottom depict weekly Availability, Outages, 
     Downtime, and Failure Clustering (see \S\ref{s:corr}), 
     respectively. The X axis in all plots is week. The Y axis in 
     Downtime row is CPU-hours and in Failure Clustering row, the number
     of machines/nodes involved.}
    \label{fig:DataCollage}
   \end{center}
\end{figure*}

Table~\ref{tbl:failuredataO2K} and ~\ref{tbl:failuredataPT} 
and Figure~\ref{fig:DataCollage} summarize the failure data 
for the three HPC systems. There are two kind of availability 
measures. The usual availability is computed as
\[
1 - \frac{\sum (\mbox{\# Down CPU} \times \mbox{Downtime})}{\mbox{\# Total CPU} \times \mbox{Total time}}
\]
The scheduled availability (S Avail) removes the 
Scheduled Maintenance downtime from consideration
and only counts scheduled uptime as total time, so
it is computed as
\[
1 - \frac{\sum (\mbox{\# Down CPU} \times \mbox{Unsched. Downtime})}{\mbox{\# Total CPU} \times \mbox{Sched. time}}
\]
Note that in O2K's case, the twelve machines have different 
number of CPUs, so ``\# Down CPU'' is the number of CPUs on the 
failed machine. In Platinum and Titan's case, the 
``\# Down CPU'' is 2. 

For the whole system of O2K, the TBF reported in 
Table~\ref{tbl:failuredataO2K} is actually TBF, and the 
downtime is the weighted average of individual machine 
downtimes:
\[
\frac{\sum ( \mbox{\# Down CPU} \times \mbox{Downtime} ) }{\mbox{\# Total CPU}}
\]

From the data it is obvious that software halts account for most 
outages (59-83\%), but the average downtime (i.e. MTTR) is only 
0.6-1.5 hours. On the other hand, although the fraction of hardware 
outages is meager (1-13\%), average hardware downtime is 
the greatest among all unscheduled outage types (6.3-100.7 hours). 
This is reasonable because hardware problems usually 
requires ordering and replacing parts and performing tests, while
many software problems can be fixed by reboot.

We contacted the NCSA staff about the hardware failure causes of
PC clusters.
We were told that there were two or three cases where power supplies
needed to be replaced; otherwise, the main cause of hardware
outages is the Myrinet, including network cards, cables, and switch cards.
A network card resides at a host PC and is connected by cables to
the Myrinet switch enclosure. A Myrinet switch enclosure stacks
many Myrinet M3-SPINE switch cards. The usual symptom that prompts a
network card or switch card replacement is there are excessive CRC 
check errors. Sometimes the self-testing in a switch card may fail 
and lead to replacement. Cable replacements also occurred because 
the ``ping'' query packets cannot get through. 

The availability is lower for O2K because when one of its machine 
is down, as much as one-sixth of the overall system capacity could 
disappear (e.g. machine B, which has 256 CPUs.) This is unlike PC 
clusters in which each node usually contains no more than 8 CPUs, 
so the availability could degrade more gracefully, assuming the 
outage is not catastrophic such as a power failure or network 
partitioning. Although monolithic single-system-image machines 
benefit from ease of administration, a unified view of process 
space, and extremely fast interprocess communication, it seems 
large systems composed of finer-grained management units are more 
favorable in terms of availability.

For O2K, the machine-wise TBFs and TTRs are skewed toward small 
values. Eleven of twelve machines have MTBF 
greater than 8 days, but the medians of TBF are mostly smaller than
 4 days. For TTR, nine machines' MTTR are greater than 2.5 hours, yet 
the medians are 0.3-0.9 hours. The same phenomenon also occurs 
on Platinum and Titan's node TTR. These prompt us to study 
examine closely the distributions of TBF and TTR, which we
documented our findings in the next section.

\Section{Failure Distribution}
\label{s:dist}
In analytical modeling, the distributions of TBF and TTR
are key components for obtaining precise results \cite{TrivediBook} 
because distributions of the same mean and variance can still 
yield very different outcomes. In this section, we investigate the 
distributions of TBF and TTR with the assumption that 
failures and repairs are all independent.

We first choose a set of distributions as our parametric probability 
models and seek the parameters that best fit the data to 
these models. An open-source statistical package called WAFO 
\cite{WafoPackage} is used to find parameters. Then we apply 
chi-square test as goodness-of-fit test to pick the best-fit 
distribution.

Our selection of probability models includes exponential, gamma
distribution and a family of heavy-tail distributions (Weibull, 
Truncated Weibull, Log-normal, Inverse normal, and Pareto 
\cite{JohnsonBook}). Heavy-tail means the complementary cumulative 
distribution function  $1-F(x)$ decays more slowly than exponentially. 
Heavy-tail distributions are chosen because many failure data 
studies (e.g. \cite{Nurmi:03,Heath:01}) have shown that they are actually 
more prevalent than exponential distribution, which is 
commonly assumed in probability models to make analysis 
tractable.

\begin{figure}
   \begin{center}
   \includegraphics{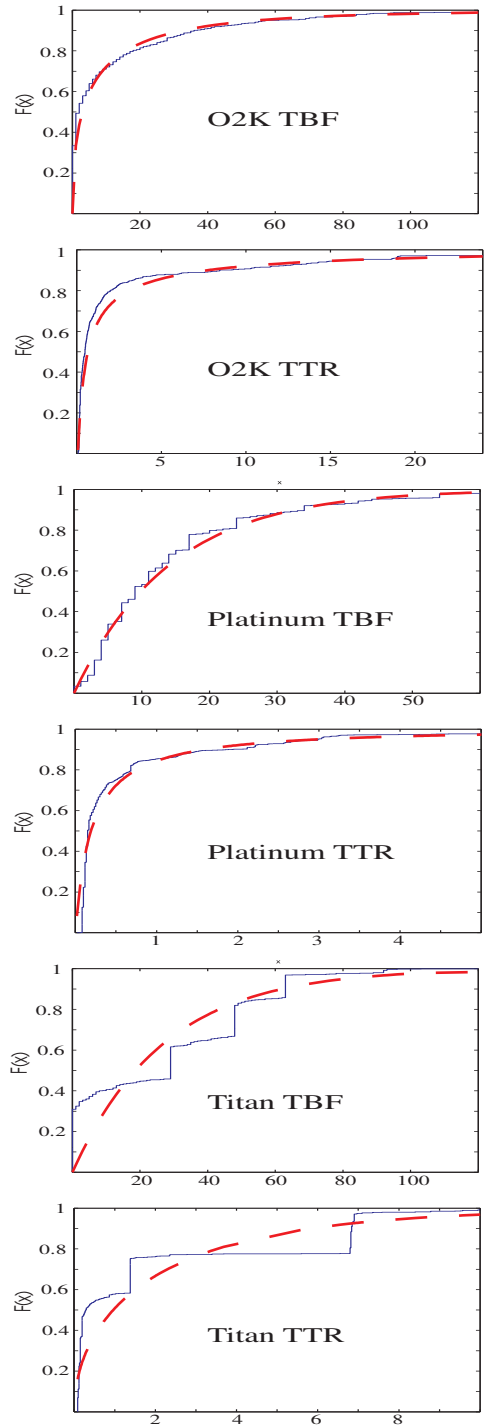} 
    \caption{Distributions of node TBF and TTR. Dashed line is 
    the fitting distribution.     The X axis in TBF plots is 
    day and in TTR plots, hour.}
    \label{fig:DistCollage}
   \end{center}
\end{figure}

For each system, we conglomerate TBF and TTR data of all 
machines/nodes and present their distributions and fitting 
functions in Figure~\ref{fig:DistCollage}. For O2K, the TTR is 
fit by Inverse normal 
$f(x) = 1.87(2 \pi x^3)^{-0.5} \exp(-12.76 (x-0.37)^2/x) $
and TBF by Weibull $F(x) = 1 -  \exp( -5.61 x^{0.5})$.
For Platinum, the TTR is fit by Truncated Weibull
$F(x) = 1 - \exp (- 6.79 (x+0.14)^{0.15} +5.07)$
and TBF by Exponential $F(x) = 1 - e^{-0.07x}$.
For Titan, the TTR is fit by Gamma 
$f(x) = 0.27 x^{-0.51} e^{-0.23x}$
and TTR by Exponential $F(x) = 1 - e^{-0.037x}$.

The distributions of Titan's failure data have 
staircase-like shapes unfound in other 
two systems'. For example, there are two sudden 
shoot-ups at 1.4 hour and 6.8 hour in Titan's TTR 
distribution. The shoot-ups mean that there were 
massive nodes down for about the same period of 
time, which implies a possibility of correlated 
failure. To understand this anomaly, we perform
a failure correlation analysis, as described in
the next section.

\Section{Failure Correlation}
\label{s:corr}
In the last section we assumed the failures are
independent and derived the failure distribuion.
Failure independence is a common assumption in 
reliability engineering to simplify analysis
and system design. However, many statistical tests
and log analyses showed that real-world distributed 
computing environments do exhibit correlated 
failures. 

In this section, we investigate how outages 
of different machines relate to each other by 
clustering approach \cite{NTfailureAnalysis1}. Roughly 
speaking, this approach groups failures which are 
close either in space or in time. It should be 
emphasized that the correlation resulted from 
clustering is purely statistical and does not 
imply the failures really have cause-and-effect 
(causal) relationship. Since our collection of 
failure log lacks error details, we can only
rely on statistics to find correlation.

To not confuse with the word ``cluster'' in ``PC 
clusters,'' we will refer to a failure cluster as 
a ``batch.'' We define a batch to be a time 
period $[T_1,T_2]$ in which every day there is 
at least one outage (regardless of type), and 
no outages occur on day $T_1-1$ or $T_2+1$. 
Put another way, we coalesce into a batch the 
failures of different machines/nodes that occur 
in consecutive days. The bottom row of
Figure~\ref{fig:DataCollage} illustrates the 
results. The width and height of a rectangle
indicate the duration and the machine/node count
of that batch, respectively.

Using this method, we found there are 79 batches 
for O2K, accounting for 55 percent of all outages. 
Eight-five percent of batches last for no more than 
three days, and 89 percent of batches involve no
more than four machines. There are four batches
that involve all twelve machines. In week 31, the
failure was caused by power or air conditioning
problem and was followed a two-day maintenance.
In week 35, there was a system-wide maintenance 
on the first day, but some machines experienced 
hardware halts and all were again taken offline for 
maintenance on the second day, and all machines 
had short software problems on the last day. In week
78, a system maintenance occurred and lasted 37-91 
hours. The last catastrophic outage occurred on 
week 97 due to power problems. Note that the massive
outages in week 31,35, and 78 are also reflected
as spikes in Availability and Downtime plots.

The failure clustering plot also reveals some 
possible failure correlation in Platinum and 
Titan systems. Statistically speaking, the chance
of a batch having a great deal of outages in a 
short time (e.g. the razor-thin rectangles in 
the bottom row of Figure~\ref{fig:DataCollage}) is 
close to zero. Thus, a reasonable explanation for such 
an occurrence is failure correlation. To justify this 
claim, we take Platinum system as an example. There is 
a batch in week 4 which contains 501 nodes in one day. 
If we assumes failures are independent and TBF 
has exponential distribution, then the number of
failures in a given duration follows Poisson distribution.
So the chance of at least 501 outages in 
one day is 
$\sum_{n=501}^{\infty} (e^{-30} 30^{n}/n!) = 6.3 \times 10^{-14}$
where $30$ is the average number of outages per day
of Platinum system. After checking the log, it
shows that particular outage is Software Halt and gives
5-15 minutes downtime.

Titan system's failure correlation is even more
conspicuous. The three peaks represent massive
outages at week 10, due to a 9 minute software halt
followed by 6.8 hours of hardware halt, at week
14, due to 1.4 hours of power failure, and at week
21, due to 6.8 hours of power failure. The 1.4 and
6.8 hours of downtime explains the two sudden rises
in Titan's TTR distribution in Figure~\ref{fig:DistCollage}
as most nodes experienced them.
The three staircases in Titan's TBF distribution reflect
the intervals among the three massive outages, which
are 64, 29, and 48 days. As in O2K's case, the three
outages of Titan are also mirrored in Availability, 
Outages, and Downtime plots.

\Section{Related Work}
\label{s:rel}
Field failure data analysis of very large HPC systems
is usually for internal circulation and is almost never 
published in detail. Nevertheless, there are several talks 
and reports that shed light on the administration experience
of some of the world's most powerful supercomputers. 

Koch \cite{Koch:02} reported the situtaiton of ASCI White.
A whole-system reboot of ASCI White takes 4 hours and 
preventive maintenance is performed weekly, with separate 
periods for software and hardware. Machine problems occurred 
in every aspect of the system. Transient CPU faults 
generated invalid floating-point numbers, and it took 
great effort to spot these corrupted nodes because they passed 
standard diagnostic tests and only failed in real programs. 
Bad optical interconnects led to non-repeatable link errors 
which corrupted the computation because these errors could 
sneak through network host firmware without being detected. 
The storage system was not 100\% dependable either. 
The parallel file-system sometimes failed to return I/O error 
to the user program when the program was dumping restart files. 
In addition, the archival subsystem's buggy firmware corrupted 
restart files and made the user program fail to launch. 

Seager \cite{Seager:03} showed that the reliability of the 
ASCI White improved over time as MTBF increased steadily from 
as short as 5 hours in January 2001 to 40 hours in February 
2003. Except uncategorized failures, the storage system 
(both local disks and IBM Serial Disk System) 
is the main source of hardware problems. Next to disks is 
CPU and third-party hardware troubles. For software, 
communication libraries and operating systems contributed
the most interruptions.

Morrison \cite{Morrison:03} reported operations of the
ASCI Q during June 2002 thru February 2003. The MTBI 
(mean time between interruption) is 6.5 hours, and on 
the average there were 114 unplanned outages per month. 
Putting storage subsystem aside, hardware problems account 
for 73.6\% of node outages, with CPU and memory modules 
being responsible for over 96\% of all hardware faults 
(CPU is 62.5\% and memory is 33.6\%.) Network adaptors or 
system boards seldom fail. 

Levine \cite{Levine:03} described the failure statistics of 
Pittsburgh Supercomputing Center's supercomputer Lemieux: MTBI 
during April 2002 to February 2003 is 9.7 hours, shorter than 
predicted 12 hours. The availability is 98.33\% during 
mid-November 2002 to early February 2003.

The National Energy Research Scientific Computing Center 
(NERSC) houses several supercomputers and their operations 
are summarized in NERSC's annual self-evaluation reports 
\cite{NERSC02}. During August 2002 to July 2003, their 
largest supercomputer Seaborg reached 98.74\% scheduled 
availability, 14 days MTBI, and 3.3 hours MTTR. Storage 
and file servers had similar availability. Two-thirds 
of Seaborg's outages and over 85\% of storage system's 
outages are due to software.

\Section{Conclusions}
\label{s:concl}
In this paper we reported the failure data analysis of
three NCSA HPC systems, one of which is an array of
distributed shared memory mainframes and the rest are
PC clusters. The results show that the availability 
is 98.7-99.8\%. Most outages are due to software halts, 
but the downtime per outage is highest due to 
hardware halts or scheduled maintenance.
We also sought the distributions of time-between-failures
and time-to-repairs and found some of them exhibit 
heavy-tail distributions instead of exponential. 
Finally, we applied failure clustering analysis and identified
several correlated failures. Because failure data
analysis of HPC system is scarce, we believe this paper 
provides very valuable information for researchers and 
practioners working on reliability modeling and engineering.

\section*{Acknowledgements}
Special thanks go to Nancy Rudins and Brian Kucic of NCSA
for providing the failure data.

\bibliographystyle{latex8}
\bibliography{failure2}

\end{document}